\documentstyle[prl,aps,multicol,epsfig]{revtex}

\begin{document}
\large

\title{Search for optimum labeling schemes in qubit systems for Quantum Information processing by NMR}
\author{Ranabir Das$^1$, 
Sukhendu Chakraborty $^2$\footnote{ {\small \it Winter trainee from Department of Computer Science, Indian Institute of Technology, Roorkee, India}}, 
K. Rukmani$^1$\footnote{ {\small \it Summer teacher fellow from Department of Physics, University of Bangalore, Bangalore, India}}
 and Anil Kumar $^{1,2,}$\footnote{ {\small \it DAE-BRNS Senior Scientist}\\}\\
        $^1$ {\small \it  Department of Physics, Indian Institute of Science, Bangalore, India}\\
        $^2$ {\small \it Sophisticated Instruments Facility, Indian Institute of Science, Bangalore, India}\\}

\maketitle

\begin{abstract}
Optimal labeling schemes lead to efficient experimental protocols for quantum information processing by nuclear magnetic 
resonance (NMR). A systematic approach of finding optimal labeling schemes for a given computation is described here. The scheme is 
described for both quadrupolar systems and spin-1/2 systems.
Finally, one of the optimal labeling scheme has been used to experimentally 
implement a quantum full-adder in a 4-qubit system by NMR, using the technique of transition selective pulses.
\end{abstract}
\section{Introduction}
 In future quantum computers may solve certain problems which are intractable by the classical computers \cite{db,ic}. Several 
quantum algorithms have been devised which use the quantum-mechanical
 properties of the physical systems to solve problems with more speed and less space \cite{pw,deujoz,gr}. 
Implementation of the quantum algorithm requires a coherent control over the 
physical systems which are used for computation. In recent trends, a great 
deal of emphasis is laid on how to simplify the experimental scheme, so as to keep coherent 
control and avoid errors \cite{cont,cont1,cont2}. Among various techniques, 
nuclear magnetic resonance (NMR) has emerged as a suitable system for demonstrating  
quantum information processing with small number of qubits \cite{na,dg,ernst,djchu,djjo,kd,lo,ci,nat,ranatomo,ranapra1}.
 In liquid-state NMR, the information processing is carried out by the use of spin/qubit-selective pulses
 intermittent with evolutions under the system Hamiltonian \cite{ic,dg}. Another experimental technique is to use 
transition selective pulses along with qubit selective pulses \cite{kd,ne,mur,ron}. Transition selective 
pulses are tuned to the resonance frequency of a single-quantum transition causing irradiation 
at a single line of the spectrum while keeping the others unperturbed. A transition 
selective $\pi$ pulse tuned at a specific transition exchanges the amplitudes between 
the two connecting states. This fact can be used to simplify the 
implementation of several logical operations. For example,  in a N-qubit system 
a $C^{N-1}NOT$ gate would require a complex pulse sequence with a series of 
qubit selective pulses inter spaced with Hamiltonian evolutions; where as it requires only one 
transition selective $\pi$ pulse between the states $\vert 111..10 \rangle$ and $\vert 111..11 \rangle$.

Further, it has been shown earlier, that relabeling of states simplifies the experimental 
protocol of certain operations \cite{mur,pramana}. While implementing half-adder and subtracter
 operations in a quadrupolar system, relabeling led to an efficient 
experimental scheme which requires less number of pulses than the conventional labeling \cite{mur}.
The idea behind relabeling is as follows: For spin-1/2 systems, conventional labeling (CL) 
uses the following logic. The state in which all the spins are in identical state, such as 
$\vert \alpha \alpha \alpha ...\alpha \rangle$ are labeled as $\vert 000...0 \rangle$ and each 
spin flip is labeled as bit flip, namely $\vert \alpha \beta \alpha ...\alpha \rangle$=
$\vert 010...0\rangle$. This logic labels each state with a well identified label and 
 leads to identification of a spin as a qubit. Spin-selective pulses then act as 
qubit selective pulses and many pulse schemes have been developed which use spin (or qubit)
 selective pulses inter missioned with Hamiltonian (or exchange coupling, J) evolution periods \cite{ic,na,dg}. 
On the other hand quantum information processing (QIP) has also been demonstrated using spins $>$1/2 
nuclei using quadrupolar couplings in molecules partially oriented in liquid crystalline media. 
In such systems, spin is no more a qubit and it has been demonstrated that the $2^N$ energy levels 
of such systems can be utilized as N qubit systems. So far only spin-3/2 and 7/2 systems have been 
utilized respectively yielding 2 and 3 qubit systems \cite{fun,mulf,sim,ne,mur,pramana,ranapra}.
 In such systems a bit flip is not a spin-flip 
while it can be treated as a qubit flip. One can follow a CL scheme in which the lowest (or highest) 
energy level can be given the label $\vert 000 \rangle$ and each subsequent level can be labeled in 
increasing order of binary numbers (CL) or single bit flips (Gray code) as shown in Table 1.
 It has been conjectured earlier that all such schemes are acceptable so long as a single label 
is attached to one level and the scheme is retained throughout a given set of computations \cite{mur,pramana}. Indeed 
it was demonstrated that it is acceptable to search for "optimum labeling" scheme (OLS) such that 
a minimum number of unitary transforms are needed for a given set of computations \cite{mur}. 

 The utility of OLS is explained in the following: A transition-selective pulse
 has low power and small bandwidth and it excites a selected single quantum transition.
That is, it can cause an operation between two states which
differ by $\Delta m= \pm1$, where $m$ is the magnetic quantum number. Suppose an
 operation requires a transformation between two states which differ by $\Delta m \neq \pm1$.
Then a single transition-selective $\pi$ pulse will not suffice. The CL scheme then looks for some intermediate states
which are connected between themselves as well as to these two states by $\Delta m= \pm1$ transitions, and applies a sequence of 
several transition selective $\pi$ pulses, whereby the two states are transformed
 via the intermediate states. However one can always relabel the energy levels such that
these two states are levels which are connected by $\Delta m= \pm1$ transition. A single transition selective 
$\pi$ pulse would then suffice. With this logic one can find an optimal labeling which reduces the computation to a minimum number of 
such transition selective pulses.   

 The optimum labeling scheme for half adder and subtracter using an oriented spin-7/2 (3-qubit) system 
given in column C of table 1 was arrived 
by trial and and error \cite{mur}. However, by no means this scheme is unique and there must be many more labeling 
schemes with equal efficiency. Furthermore for higher qubit systems the trial-and-error method will 
become laborious and inefficient. Therefore there is a need for a systematic approach to 
this problem of finding optimum labeling scheme for a given computation or a set of computations. 
This paper deals with one such approach. Section II outlines a protocol to search for 
optimum labeling schemes, section III introduces full adder operation, section IV gives the approach of search for 
optimum labeling schemes in case of multiple operations with an example of a full adder + swap2,4, 
and section V contains an experimental implementation of a full adder by a 4-qubit weakly coupled spin-1/2 system.   

\section{Optimum labeling scheme}      
 To search for optimum labeling schemes (OLS) we start with a truth table of a computation. 
Table 2 contains a particular truth table for a 4-qubit system.
At this moment it is not important to know the logical operation this truth table represents. It is only to illustrate the procedure.   
 We search for OLS with the approach of set-theory. We consider that all the states of the system constitute a universal set $\{S\}$.
Then from the truth table we construct maximal sets $\{S_i\}$ which are mutually exclusive subsets of $\{S\}$. To construct maximal sets,
 first input state is taken up and put in the first set $\{S_1\}$. The corresponding output is noted and is added to $\{S_1\}$
if it is not already included in it (Table 3). This process continues till all the states generated by each state in $\{S_1\}$ have been
included. In the present case, the $\{S_1\}$ set contains only one element ${\vert 0000 \rangle}$ since it transform into itself. 
 Similarly, sets $\{S_2\}$, $\{S_3\}$ and  $\{S_4\}$ contain one element each.
 The set $\{S_5\}$ is formed by noting that $\vert 0100 \rangle$
transforms to $\vert 0110 \rangle$, which in turn transforms to $\vert 0101 \rangle$, which transforms to $\vert 0111 \rangle$ which 
then transforms to $\vert 0100 \rangle$, completing the set.  
 This process is carried out for all $\{S_i\}$, by selecting a input state not forming a part of all previous $\{S_i\}$'s. This process is
 continued  till  all the states are included in exactly one of the maximal sets $\{S_i\}$ (Table 3). It is evident that the maximal 
sets  $\{S_i\}$ will be mutually exclusive. 
 
   An optimum labeling 
scheme for executing the logical operation of Table 2 by single quantum transition selective pulses is obtained by arranging the 
labels of levels in the same order as in column 1 of table 3.  
 The number of pulses for any set $\{S_i\}$ will be $\vert S_i \vert -1$, where $\vert S_i \vert $ is the 
cardinality (number of elements) of the set.
Thus the minimum number of pulses required for the execution of the logical operation of Table 2 is
\begin{eqnarray}
 N_p=\sum ^{M}_{i=1} (\vert S_i\vert -1),
\end{eqnarray}
 where M is the number of maximal sets. In the present example the number of transition selective pulses needed are 3+3+1+1=8.
 It may be pointed out that implementation of this operation in a quadrupolar system using conventional 
labeling and gray code would respectively require  12 and 10 transition selective pulses. 

    After creating the maximal sets, one has to consider
only those sets that have more than one element, as they are the ones which would require 'pulses'.
 We have seen that in each of these maximal sets, the transformations between different states
takes place in a  chain. These  chain of states should be mapped to a chain of energy levels where each
level in the chain should be connected to its previous and next level by single quantum transitions.
 Mapping the sets to the subspace of energy levels, should start with the mapping of a set having the
largest no. of elements (i.e. max. cardinality) of all the sets available and move in a decreasing order.
This mapping follows different strategies for quadrupolar systems and coupled spin-1/2 systems. 
These are outlined in the following.

\subsection{Optimum labeling for quadrupolar systems}
 The Hamiltonian of a quadrupolar nucleus partially oriented in liquid crystalline matrix, in the presence
of a large magnetic field $B_0$ and  having a first-order quadrupolar coupling is given by \cite{khel}
\begin{eqnarray}
\mathcal{H}=\mathcal{H} _Z+\mathcal{H}_Q &=&-\omega_0 I_z + \frac{e^2qQ}{4I(2I-1)} (3I^2_z-I^2)S \nonumber \\
   &=& - \omega_0 I_z+\Lambda (3I^2_z-I^2),
\end {eqnarray}
 where $\omega_0=\gamma B_0$ is the resonance frequency, $\gamma$ being the gyro magnetic ratio,
 $S$ is the order parameter at the site of the nucleus,  $e^2qQ$ is the quadrupolar coupling and
$\Lambda = e^2qQS/(4I(2I-1))$ is the effective quadrupolar coupling. Though $e^2qQ$ is of the order of several MHz,
 a small value for the order parameter ($S$) converts the effective quadrupolar coupling `$\Lambda$' into several kHz.
 In such circumstances, 
 a spin-I nucleus  has $2I+1$ non-equispaced eigenstates and $2I$ well resolved single-quantum transitions separated by 
 effective quadrupolar coupling `$\Lambda$'. It has been demonstrated earlier, that such systems can be treated as an N-qubit 
system, provided $(2I+1)=2^N$ \cite{fun}. For example a single spin-3/2 acts as a 2-qubit system and a spin-7/2 acts as a 
3-qubit system \cite{fun,mulf,sim,ne,mur,ranapra}.

 In quadrupolar systems the energy levels are in increasing order of Zeeman energy. Each level is connected 
to two adjacent levels by single quantum transitions, except terminal levels (lowest and highest energy level) which are connected by 
only one single quantum transition to its nearest level. This puts certain restrictions in identification of maximal sets
 with the energy levels. An example of labeling scheme for the operation in Table 2 is given in Fig. 1.
 The maximal sets are shown in the energy
level diagram along with the transition selective $\pi$ pulses which are required to implement the truth table of table 2 
in this labeling scheme. Let us take the case of S$_5$ which has four states. They are being mapped in a subspace of the
energy level diagram to four energy levels which are in a chain. Then the required transformations
can be achieved by three $\pi$-pulses applied in the reverse order of the chain. Hence the pulses are to be applied in the order
($\pi_1\pi_2\pi_3$) as shown in Fig. 1, 
\begin{eqnarray}
(\pi_1)^{\vert 0101\rangle \leftrightarrow \vert 0111\rangle}_y
(\pi_2)^{\vert 0101\rangle \leftrightarrow \vert 0110\rangle}_y
(\pi_3)^{\vert 0110\rangle \leftrightarrow \vert 0100\rangle}_y &=&
\pmatrix{1&0&0&0 \cr 0&0&0&1 \cr 0&0&1&0 \cr 0&-1&0&0}
\pmatrix{1&0&0&0 \cr 0&0&1&0 \cr 0&-1&0&0 \cr 0&0&0&1}
\pmatrix{0&0&1&0 \cr 0&1&0&0 \cr -1&0&0&0 \cr 0&0&0&1}  \nonumber \\
&=& \pmatrix{0&0&1&0 \cr 0&0&0&1 \cr 0&-1&0&0 \cr 1&0&0&0}.
\end{eqnarray}
The above operator is for a subsystem of the last two qubits in the four-qubit system,
where the first two qubits are in the fixed state $\vert 01\rangle$.
S$_6$ also has a similar chain which is then mapped to a chain of levels as shown in Fig 1.
S$_7$ has a chain of two states and it can be mapped on to any two adjacent energy levels of the system. $S_8$
follows same logic. The energy levels corresponding to states of S$_7$ and S$_8$ in our labeling scheme can be seen in Fig. 1.
 However, the relabeling scheme of Fig. 1 is not unique, and many 
optimum relabeling schemes are possible.

  The basic idea of relabeling is that the $2^N$ eigenstates of a N-qubit system can be given various desired labels. 
It turns out that if the only condition is that one labels is attached to each level, then there are $2^N$! possibilities.
However, only a few of these are optimal. In the case of quadrupolar system all the energy levels differ in their 
energy by at least one Larmor frequency and  there are $2^N-1$ single-quantum transitions, for a N-qubit systems.
 Thus the number of different OLS possible in quadrupolar system is only 
a permutation of the different maximals sets,  multiplied by the number of ways the elements in each set can be arranged. 
However we note that for sets with more than one element, optimal labeling demands that the order of states must be same as that 
of the transformations; thereby allowing only two options of arranging in ascending or descending order. Hence the total number 
of optimal labeling schemes is:
\begin{eqnarray} 
P=M!~~2^k,
\end{eqnarray}
 where k is the number of maximal sets with more than one state. We note that for the example of Table 1, the number of  OLS 
 are 8!$\times 2^4$=645120  out of a total of 16!$\cong 2\times 10^{13}$ possible labeling schemes.

\subsection{Optimum labeling for spin-1/2 systems} 
 In a large magnetic field $B_0$, the energy level of a spin-1/2 nucleus is split into two by Zeeman interaction. These 
two energy levels can be labeled as $\vert 0 \rangle$ and $\vert 1 \rangle$ and hence a spin-1/2 nucleus acts as a qubit. 
 N such nuclei, having different Larmour frequencies and coupled to each other by indirect spin-spin interaction,
 constitute a N-qubit system.     
 The Hamiltonian for such a system is given by \cite{ernst},
\begin{eqnarray}
{\mathcal H}&=&{\mathcal H}_Z+{\mathcal H}_J \nonumber \\
            &=& \sum_i \omega_i I_{iz} +\sum_{i,j (i<j)} 2\pi J_{ij}\vec{I_i}.\vec{I_j}
\end{eqnarray}
 where ${\mathcal H}_Z$ is the Zeeman Hamiltonian, ${\mathcal H}_J$ is the coupling Hamiltonian, 
$\omega_i=\gamma_iB_0$ is the resonance frequency of the i$^{th}$ spin, and $J_{ij}$ is the coupling between i$^{th}$ and j$^{th}$ spin.
When $2\pi J_{ij} \ll \vert \omega_i-\omega_j \vert$, the system is said to be weakly coupled,
and the Hamiltonian can be approximated to  \cite{ernst},
\begin{eqnarray}
{\mathcal H}= \sum_i \omega_i I_{iz} +\sum_{i,j (i<j)} 2\pi J_{ij}I_{iz}I_{jz}
\end{eqnarray}
Under the approximation of Eq.(6), product of states of individual spins are eigenstates of the system and a spin can be
 treated as a qubit \cite{na,dg}. In this paper we restrict to such systems. In such cases of  N-qubit spin-1/2 systems, each 
level is connected to N other levels 
by single quantum transitions, amounting to a total of $N\cdot 2^{N-1}$ single quantum transitions. 
Hence the number of possible optimal labeling schemes is much larger than the quadrupolar systems described above, and it turns out 
that in many cases the conventional labeling scheme may be an optimum labeling scheme with some minor modifications.
 For example, for the truth table 
of Table 2, a conventional labeling scheme and the pulses required 
for this scheme are shown in Fig 2(a). Note that in the maximal set S$_5$ the transformations require three pulses,
$\pi_1$, $\pi_2$ followed by $\pi_3$, whose operator is, 
\begin{eqnarray}
(\pi_1)^{\vert 0101\rangle \leftrightarrow \vert 0111\rangle}_y
(\pi_2)^{\vert 0101\rangle \leftrightarrow \vert 0110\rangle}_y
(\pi_3)^{\vert 0110\rangle \leftrightarrow \vert 0100\rangle}_y &=&
\pmatrix{1&0&0&0 \cr 0&0&0&1 \cr 0&0&1&0 \cr 0&-1&0&0}
\pmatrix{1&0&0&0 \cr 0&0&1&0 \cr 0&-1&0&0 \cr 0&0&0&1}
\pmatrix{0&0&1&0 \cr 0&1&0&0 \cr -1&0&0&0 \cr 0&0&0&1}  \nonumber \\
&=& \pmatrix{0&0&1&0 \cr 0&0&0&1 \cr 0&-1&0&0 \cr 1&0&0&0}.
\end{eqnarray}
The above operator is for a subsystem of the last two qubits in the four-qubit system,
where the first two qubits are in the fixed state $\vert 01\rangle$.
However, the transformation of $\pi_2$ is between 
two states $\vert 0101 \rangle$ and $\vert 0110\rangle$ which differ by 
$\Delta m=0$, and cannot be accomplished by a single transition-selective pulse. 
Hence it would seem that the experimental protocol 
would require more pulses. The number of pulses can be reduced by relabeling. Fig 2(b) shows a
 relabeled scheme where the labels of the states ($\alpha \beta \beta \alpha$ and 
$\alpha \beta \alpha \alpha$) as well as ($\beta \alpha \alpha \alpha$ and
$\beta \alpha \beta \alpha$) are interchanged so that
 $\vert 0101 \rangle$ and $\vert 0110\rangle$ and $\vert 1001 \rangle$ and $\vert 1010\rangle$ are connected by 
single quantum transitions.

  However, in this case we observe that,  by changing the sequence of pulses, one can achieve the same
transformations of S$_5$ and S$_6$ in conventional labeling scheme with minimum number of pulses. 
For example, in the maximal set S$_5$, if we change the sequence of pulses as 
$\pi_1$, $\pi_3$ followed by $\pi_2$, where $\pi_2$ is applied between the states 
$\vert 0100\rangle$ and $\vert 0101 \rangle$ (as shown in fig 2(c)), 
then the operator is the same as of Eq. (7).   
\begin{eqnarray}
(\pi_1)^{\vert 0101\rangle \leftrightarrow \vert 0111\rangle}_y
(\pi_3)^{\vert 0110\rangle \leftrightarrow \vert 0100\rangle}_y
(\pi_2)^{\vert 0100\rangle \leftrightarrow \vert 0101\rangle}_y &=&
\pmatrix{1&0&0&0 \cr 0&0&0&1 \cr 0&0&1&0 \cr 0&-1&0&0}
\pmatrix{0&0&1&0 \cr 0&1&0&0 \cr -1&0&0&0 \cr 0&0&0&1}
\pmatrix{0&1&0&0 \cr -1&0&0&0 \cr 0&0&1&0 \cr 0&0&0&1}  \nonumber \\
&=& \pmatrix{0&0&1&0 \cr 0&0&0&1 \cr 0&-1&0&0 \cr 1&0&0&0}.
\end{eqnarray}
This sequence will require only three single-quantum pulses since all the pulses are between states with $\Delta$m=$\pm$1.
Similarly, in set S$_6$; $\pi_4$, $\pi_6$ followed by $\pi_5$, will suffice (fig. 2(c)).
 In this protocol, the conventional labeling scheme requires a total of eight pulses for implementing quantum full-adder, 
which is the same as OLS and hence, conventional scheme is also optimum. 
However, for experimental convenience, relabeling may still be useful, as will be shown in the experimental section.
It may be noted from Eq. (3), (7) and (8), that the collective operator of the three transition selective pulses differ from the ideal 
operator of the transformations in S$_5$, by a controlled
phase factor \cite{dg}.  If one starts from the equilibrium state the results are identical to that of the correct full-adder. When
applied to a pure state, the phase factor must either be taken into consideration or can be corrected by adding a
controlled phase gate using transition selective z-pulses \cite{ron}.

\section{Fulladder}
 The truth table of Table 1 is actually the truth table of a quantum full-adder.
The full adder is a basic component of a conventional computer. The quantum 
full adder is also an important part of many quantum algorithms. In particular it is a key step in 
Shor's prime factorization algorithm, where it is necessary to perform modular exponentiation 
$f(x)=a^xmodM$ \cite{pw}. A classical full adder (Table 4) adds bits "A" and "B" and carry "C$_0$"  
to give a sum "S" and a carry "C". 
 This operation is not reversible. Quantum full adder however needs to be reversible, 
and hence an extra ancillary bit is added in the input "L" to make the operation reversible. 
The truth table then becomes exactly the one is given in Table 2, where X$_1$=C$_0$, X$_2$=A, X$_3$=B and X$_4$=L,  
Y$_1$=C$_0$, Y$_2$=A, Y$_3$ is the sum (=C$_0$$\oplus$A$\oplus$B=S), and 
Y$_4$ is the carry (=L$\oplus$(AB$\oplus$AC$_0$$\oplus$BC$_0$)=C$_1$). Fig.3 contains the circuit for quantum full-adder and 
Fig. 1 one of the many possible optimum labeling schemes for the quantum full adder in a 4-qubit (spin-15/2) quadrupolar system.

\section{multiple operations}
  The optimal labeling for a sequence of logical operations can be constructed in a manner which is similar to the one outlined 
in section II. 
For example, if one wishes to implement a swap operation between 2nd and 4th qubit after implementing 
full-adder, then the maximal sets have to be 
constructed from the truth table of combined operation of full-adder+swap-2,4. For full-adder+swap-2,4, the maximal sets are  
 S$_1$=$\{\vert 0000\rangle\}$, S$_2=\{\vert 0001\rangle,\vert 0100\rangle,\vert 0011\rangle,\vert 0110\rangle,\vert 0101\rangle,
\vert 0111\rangle\}$, S$_3=\{\vert 0010\rangle\}$ and S$_4=\{\vert 1000\rangle,\vert 1010\rangle,
\vert 1100\rangle,\vert 1101\rangle,\vert 1001\rangle,\vert 1110\rangle,\vert 1111\rangle,\vert 1011\rangle\}$.
For implementing full-adder+swap2,4 in a 4-qubit quadrupolar system, this labeling scheme would require 12 transition selective pulses.
Often various logical operations do not commute. For example, these two logical operations do not commute, and hence, 
 if one wants to implement the operations in the reverse order, namely swap-2,4+full-adder,
 the truth table of the combined operation is different and so is the order of elements in the maximal sets:
 S$_1$=$\{\vert 0000\rangle\}$, S$_2=\{\vert 0001\rangle,\vert 0110\rangle,\vert 0011\rangle,\vert 0101\rangle,\vert 0111\rangle,
\vert 0100\rangle\}$, S$_3=\{\vert 0010\rangle\}$ and S$_4=\{\vert 1000\rangle,\vert 1010\rangle,
\vert 1001\rangle,\vert 1101\rangle,\vert 1100\rangle,\vert 1011\rangle,\vert 1111\rangle,\vert 1110\rangle\}$. 
However, it is evident that  swap-2,4+full-adder also requires 12 transition selective pulses. 
It may be mentioned that the implementation of full-adder+swap-2,4 by
CLS would require 24, and Gray Code would require 26 transition selective pulses in a 4-qubit quadrupolar system. 
\section{Experimental}
   The molecule 2-3 diflouro 6-nitrophenol (dissolved in CDCl$_3$+1 drop D$_2$O) has 4 weakly coupled spin-1/2 
nuclei, acts as a 4 qubit system and was chosen to implement the 
quantum full-adder (Fig. 4). The proton of the phenol group is exchanged with the D$_2$O.
 The two remaining protons and the the two fluorine nuclei constitute the four qubit system. 
 The equilibrium spectrum of each nucleus  is shown in Fig. 4.  The chemical shift difference between the 
two Fluorine spins is 16 kHz while that between the two  protons is 250 Hz. The couplings range from 19.1 Hz to -2.3 Hz.
 The energy level diagram (Fig. 5) was constructed with two independent methods. In method 
(i) a  transition tickling experiment was done on 
all the 32 transitions to yield the connectivity matrix \cite{pag} from which a simple calculation constructed the 
energy-level diagram. In method (ii) a two-dimensional (2D) heteronuclear Z-COSY  experiment was  performed \cite{rang}. 
The sign of the peaks in the 2D spectrum yielded the connectivity matrix which confirmed the energy-level 
diagram obtained by the method (i).

 We start with the equilibrium state and implement the full-adder using transition selective pulses. While applying the 
selective pulses we took some factors into consideration. First, the transition selective pulses have to be tuned 
at a specific frequency with a narrow bandwidth so as to prevent the other lines from being perturbed. But again narrow 
bandwidth implies long pulses which in turn increases the experimental time during which relaxation sets in.  
 One has to optimize the experiment time so as to avoid errors due to relaxation. 
The schemes in Fig 2(b) or 2(c) show that to implement full-adder in the 4-qubit system, 
one needs pulses only on two of the four qubits. The specific transitions to be inverted are, however, far apart from each 
other with transitions between them. Hence, such a labeling would require pulsing individual transitions. 
 Pulses with high selectivity have to be applied; needing long pulse 
lengths and experimental times, and leading to significant effects of decoherence. On the other hand, by relabeling the energy levels 
suitably, the experimental protocol can be simplified. Figure 5 shows a relabeling which allows pulsing six 
transitions of one spin (I$_4$ of our system) followed by two transitions of the another spin (I$_3$ of our system). 
Moreover, these transitions were so chosen such that they are adjacent to each other in the frequency 
spectrum and can be pulsed simultaneously, as shown below.   
  
First, we applied a spin-selective $\pi$ pulse which inverted all the 8 transitions of I$_4$.
Second, we applied  another selective $\pi$ pulse on two transitions (first two from left in Fig. 4) of I$_4$. 
The frequency  of this selective pulse was tuned at the center of the two transitions and 
pulse power was adjusted to cause a $\pi$ rotation of the two transitions. Thus these two transitions get an effective rotation 
of $2\pi$ whereas the other six transitions are rotated by $\pi$. Thus the states connected by these two transitions 
 have their equilibrium populations restored, while the states connected by the other six transitions
 will have their populations interchanged.  Experimentally  this scheme is preferred  
because it allows short duration pulses and faster implementation.

 Subsequently, two pulses $\pi_3\pi_6$ were applied on two transitions of spin I$_3$ (5th and 6th from the left in
Fig. 4) as directed by our labeling scheme. After each selective pulse, a gradient was applied to kill any 
coherences created due to imperfection of r.f. pulses. Since we started with equilibrium state, our result is encoded in the 
final populations of different states. The final populations were monitored by using a non-selective small flip-angle (5$^o$) pulse,
 which maps, within linear approximation, the deviation populations into the intensities of various transitions. 
The obtained spectrum is shown in Fig. 6 with expected results shown as a stick diagram underneath for each spin.
 The results have been reproduced within $18\%$ of their expected intensities.
 The deviation from the expected intensities are due to relaxation and inhomogeneity of r.f. pulses. These spectra confirm the 
implementation of quantum full adder operation.

\section{Conclusion}
In this paper we  have outlined a protocol to find optimum labeling schemes for specific computations. While in quadrupolar systems 
OLS provides experimental schemes requiring less number of pulses, in spin-1/2 systems it helps to keep a better control 
over coherence and reduce experimental errors. This relabeling has been utilized for implementation of 
 quantum full-adder in a 4-qubit spin-1/2 system by transition selective pulses. Quantum full-adder has also been realized 
using Hamiltonian evolution by Chuang et. al. as a subroutine of Shor's algorithm \cite{nat}.  

 The search of higher qubits has led researchers to use homonuclear spin systems oriented in liquid crystalline matrices. 
In such systems, the homonuclear spins become strongly coupled and can no longer be treated as qubits \cite{er}.
Since the spins loose their identity as qubits, conventional labeling scheme is not defined
and Hamiltonian evolution method is not applicable, in these systems. It has been 
demonstrated that the $2^N$ eigenstates of N spin-1/2 strongly coupled nuclei can still be treated as a N-qubits system 
and quantum information processing can be performed using single-quantum transition-selective pulses \cite{strong}. 
In such systems, while the number of allowed single quantum transitions are more than that in weakly coupled systems, 
the number of transitions having significant (observable) intensities, in some cases, may be less \cite{khel}. 
In such cases OLS can be used to optimally label the eigenstates and 
perform computations utilizing observable single quantum transitions \cite{rang}. 

\section{Acknowledgement}
The authors thank T. S. Mahesh and  Rangeet Bhattacharya for useful discussions. The
 use of DRX-500 NMR spectrometer funded by the Department of
Science and Technology, New Delhi, at the Sophisticated Instruments Facility, Indian Institute of Science, 
Bangalore, is also gratefully acknowledged. K. Rukmani  is grateful to 
Indian Academy of Science, Bangalore for a summer teacher fellowship at Indian Institute of Science, Bangalore. 
AK acknowledges "DAE-BRNS" for the award of "Senior Scientists scheme".

\pagebreak

\hspace*{9cm}
Table 1 \\
Conventional labeling, Gray code and Optimum labeling (for half-adder and subtracter operation) in a spin-7/2 system.\\\\
\hspace*{5cm}
\begin{tabular}{|c|c|c|c|} \hline
~~~Energy level~~~~&~~~~A~~~~&~~~~B~~~~&~~~~C~~~~ \\
~~~m~~~~&~~~~CL~~~~&~~~~GRAY~~~~&~~~~Optimum~~~~ \\ \hline
~~~7/2~~~~&~~~~000~~~~&~~~~000~~~~&~~~~000~~~~ \\
~~~5/2~~~~&~~~~001~~~~&~~~~001~~~~&~~~~010~~~~ \\
~~~3/2~~~~&~~~~010~~~~&~~~~011~~~~&~~~~011~~~~ \\
~~~1/2~~~~&~~~~011~~~~&~~~~010~~~~&~~~~001~~~~ \\
~~~-1/2~~~~&~~~~100~~~~&~~~~110~~~~&~~~~101~~~~ \\
~~~-3/2~~~~&~~~~101~~~~&~~~~100~~~~&~~~~110~~~~ \\
~~~-5/2~~~~&~~~~110~~~~&~~~~101~~~~&~~~~111~~~~ \\
~~~-7/2~~~~&~~~~111~~~~&~~~~111~~~~&~~~~100~~~~\\
\hline
\end{tabular}
\pagebreak

\hspace{6cm}
Table 2 \\
\hspace*{3cm} Truth table for a certain logical operation.\\\\
\hspace*{3.8cm}  INPUT \hspace*{2.5cm} OUTPUT \\
\hspace*{3cm}
\begin{tabular}{|cccc|cccc|} \hline
~~~X$_1$&X$_2$&X$_3$&X$_4$~~~~~~~&~~~~~~~~Y$_1$&Y$_2$&Y$_3$&Y$_4$~~~~ \\ \hline
~~~0&0&0&0~~~~~~~~&~~~~~~~~0&0&0&0~~~~ \\
~~~0&0&0&1~~~~~~~~&~~~~~~~~0&0&0&1~~~~ \\
~~~0&0&1&0~~~~~~~~&~~~~~~~~0&0&1&0~~~~ \\
~~~0&0&1&1~~~~~~~~&~~~~~~~~0&0&1&1~~~~ \\
~~~0&1&0&0~~~~~~~~&~~~~~~~~0&1&1&0~~~~ \\
~~~0&1&0&1~~~~~~~~&~~~~~~~~0&1&1&1~~~~ \\
~~~0&1&1&0~~~~~~~~&~~~~~~~~0&1&0&1~~~~ \\
~~~0&1&1&1~~~~~~~~&~~~~~~~~0&1&0&0~~~~ \\
~~~1&0&0&0~~~~~~~~&~~~~~~~~1&0&1&0~~~~ \\
~~~1&0&0&1~~~~~~~~&~~~~~~~~1&0&1&1~~~~ \\
~~~1&0&1&0~~~~~~~~&~~~~~~~~1&0&0&1~~~~ \\
~~~1&0&1&1~~~~~~~~&~~~~~~~~1&0&0&0~~~~ \\
~~~1&1&0&0~~~~~~~~&~~~~~~~~1&1&0&1~~~~ \\
~~~1&1&0&1~~~~~~~~&~~~~~~~~1&1&0&0~~~~ \\
~~~1&1&1&0~~~~~~~~&~~~~~~~~1&1&1&1~~~~ \\
~~~1&1&1&1~~~~~~~~&~~~~~~~~1&1&1&0~~~~ \\
\hline
\end{tabular}
\pagebreak

\hspace{7cm}
Table 3 \\
\hspace*{3cm} Construction of Maximal sets for the operation of Table 2.\\\\
\begin{tabular}{|cc|} \hline
Chains & Maximal sets \\ \hline
$\vert 0000\rangle$ & S$_1=\{\vert 0000\rangle\}$ \\
$\vert 0001\rangle$ & S$_2=\{\vert 0001\rangle\}$ \\
$\vert 0010\rangle$ & S$_3=\{\vert 0010\rangle\}$ \\
$\vert 0011\rangle$ & S$_4=\{\vert 0011\rangle\}$ \\
$\vert 0100\rangle \rightarrow \vert 0110\rangle \rightarrow \vert 0101\rangle \rightarrow \vert 0111\rangle
\rightarrow \vert 0100\rangle$~~~~
&~~~~ S$_5=\{\vert 0100\rangle,\vert 0110\rangle,\vert 0101\rangle,\vert 0111\rangle \}$ \\
$\vert 1000\rangle \rightarrow \vert 1010\rangle \rightarrow \vert 1001\rangle \rightarrow \vert 1011\rangle
\rightarrow \vert 1000\rangle$~~~~
&~~~~ S$_6=\{\vert 1000\rangle,\vert 1010\rangle,\vert 1001\rangle,\vert 1011\rangle \}$ \\
$\vert 1100\rangle \rightarrow \vert 1101\rangle \rightarrow \vert 1100\rangle$ & S$_7=\{\vert 1100\rangle,\vert 1101\rangle \}$ \\
$\vert 1110\rangle \rightarrow \vert 1111\rangle \rightarrow \vert 1110\rangle$ & S$_8=\{\vert 1110\rangle,\vert 1111\rangle \}$ \\ \hline
\end{tabular}

\pagebreak
\hspace*{5cm}
Table 4 \\
\hspace*{3cm} Truth table of a classical Full-Adder\\\\
\hspace*{3cm}
\begin{tabular}{|ccc|cc|} \hline
~~C$_0$~~~~&~~~~A~~~~&~~~~B~~~~&~~~~S~~~~&~~~~C$_1$~~~~ \\ \hline
~~0~~~~&~~~~0~~~~&~~~~0~~~~&~~~~0~~~~&~~~~0~~~~ \\
~~0~~~~&~~~~0~~~~&~~~~1~~~~&~~~~1~~~~&~~~~0~~~~ \\
~~0~~~~&~~~~1~~~~&~~~~0~~~~&~~~~1~~~~&~~~~0~~~~ \\
~~0~~~~&~~~~1~~~~&~~~~1~~~~&~~~~0~~~~&~~~~1~~~~ \\
~~1~~~~&~~~~0~~~~&~~~~0~~~~&~~~~1~~~~&~~~~0~~~~ \\
~~1~~~~&~~~~0~~~~&~~~~1~~~~&~~~~0~~~~&~~~~1~~~~ \\
~~1~~~~&~~~~1~~~~&~~~~0~~~~&~~~~0~~~~&~~~~1~~~~ \\
~~1~~~~&~~~~1~~~~&~~~~1~~~~&~~~~1~~~~&~~~~1~~~~ \\
\hline
\end{tabular}
 \\

\pagebreak
{\Large Figure Captions}\\

Fig. 1. The Zeeman energy levels of a spin-15/2 nucleus along with one of the optimum labeling schemes for a 4-qubit system,  
for the logical operation given in Table 2 in a quadrupolar system. The magnetic quantum number (m), corresponding to 
each eigenstate is given in the left hand side, and the qubit labeling is given in the right hand side.  
 The maximal sets of Table 3 are shown in the energy
level diagram along with the transition selective $\pi$ pulses required to implement the truth table of Table 2
in this labeling scheme. 

Fig. 2. (a) Conventional labeling scheme and the pulses required for implementing the truth table of Table 2 in a 4-qubit 
spin-1/2 system. (b) Relabeled energy levels to implement the truth table of Table 2 with optimum pulses.
 (c) The conventional scheme of (a) with a rearrangement of the sequence of pulses to implement the logic of Table 2, with 
minimum number of pulses.

Fig. 3. Quantum circuit of a quantum full-adder operation. The two bits A and B are added with the carry from the 
previous operations, C$_0$. An ancillary bit L is included in the input to make the operation reversible. After the full-adder 
operation, the sum 
gets stored in S (S=C$_0$$\oplus$A$\oplus$B), and the carry is stored in C$_1$ (C$_1$=L$\oplus$(AB$\oplus$AC$_0$$\oplus$BC$_0$)).

Fig. 4. The molecule 2-3 diflouro 6-nitrophenol which acts as a 4 qubit system. 
 The two protons (I$_1$ and I$_3$) and the the two fluorine nuclei (I$_2$ and I$_4$) constitute the four qubit system.

 The equilibrium spectrum of each nucleus  is individually shown. The labeling of the transitions is given above each line, 
which was been determined by transition tickling and HET-zcosy experiments \cite{pag,rang}. 
Each label identifies the transition of a given spin corresponding to the states of other spins.
The chemical shift difference between the
two Fluorine spins is 16 kHz while that between the two  protons is 250 Hz. The couplings range from 19.1 Hz to -2.3 Hz.

Fig. 5. Relabeled energy level diagram for implementing a quantum full-adder in the 4-qubit system of 
2-3 diflouro 6-nitrophenol. The maximal sets were created using the scheme outlines in section II. Then the chain of 
elements in the maximals sets were mapped on to chain of states in the 4-qubit system. To simplify the experimental protocol 
it was noted that, all the pulses which commute (which also means that they are not connected to a common energy level)
can be applied simultaneously. The maximal sets are mutually exclusive and so the pulses in different sets
commute with each other. This means that $\pi_7$ and $\pi_8$ of sets S$_7$ and S$_8$ can be applied simultaneously
with the pulses of S$_5$ and S$_6$. In the set S$_5$, the labels of states $\vert 0101 \rangle$ and $ \vert 0111 \rangle$ 
were interchanged so that the pulses $\pi_1$ and $\pi_2$ can be applied simultaneously. 
Similarly, in S$_6$, by interchanging the 
states $\vert 1001 \rangle$ and $ \vert 1011 \rangle$, $\pi_4$ and $\pi_5$ can be applied simultaneously. 
Hence, by this labeling, $\pi_1\pi_2\pi_4\pi_5\pi_7\pi_8$ followed by  $\pi_3\pi_6$ would implement the quantum full-adder.
The initial (equilibrium) populations and final (after implementation of full-adder) populations 
are given beside each energy level as: Initial population(final population). 
The intensity of various transitions change after full-adder. For example, 
we note that the intensity of $\alpha \beta \beta$ transition of I$_4$ changes from +1 to -2, after
 implementation of full-adder.

Fig. 6. Implementation of the quantum full-adder in the 4-qubit system of 2-3 diflouro 6-nitrophenol. 
Starting from equilibrium, 
three selective Gaussian shaped pulses of lengths 50us, 350ms and 450ms were applied on selected transitions 
(as explained in text). Gradients were applied after each pulse to destroy any unwanted coherences created by the imperfection 
of pulses. Final populations of different states were mapped by a non-selective small flip-angle (5$^o$) pulse.   
The experimental spectrum  is shown above with the expected spectrum shown as a stick diagram underneath for 
each spin. The intensities in the stick diagram are; 0, $\pm$1 and $\pm$2 corresponding to the final populations 
of Fig. 5. All the spectra are Fourier transformed 
after multiplication of the signal with a Gaussian window function.
 The longitudinal relaxation rates for the four spins are $T^1_1=7s$, $T^2_1=3.5s$, $T^3_1=10s$ and $T^4_1=4s$.
\pagebreak
\begin{figure}
\epsfig{file=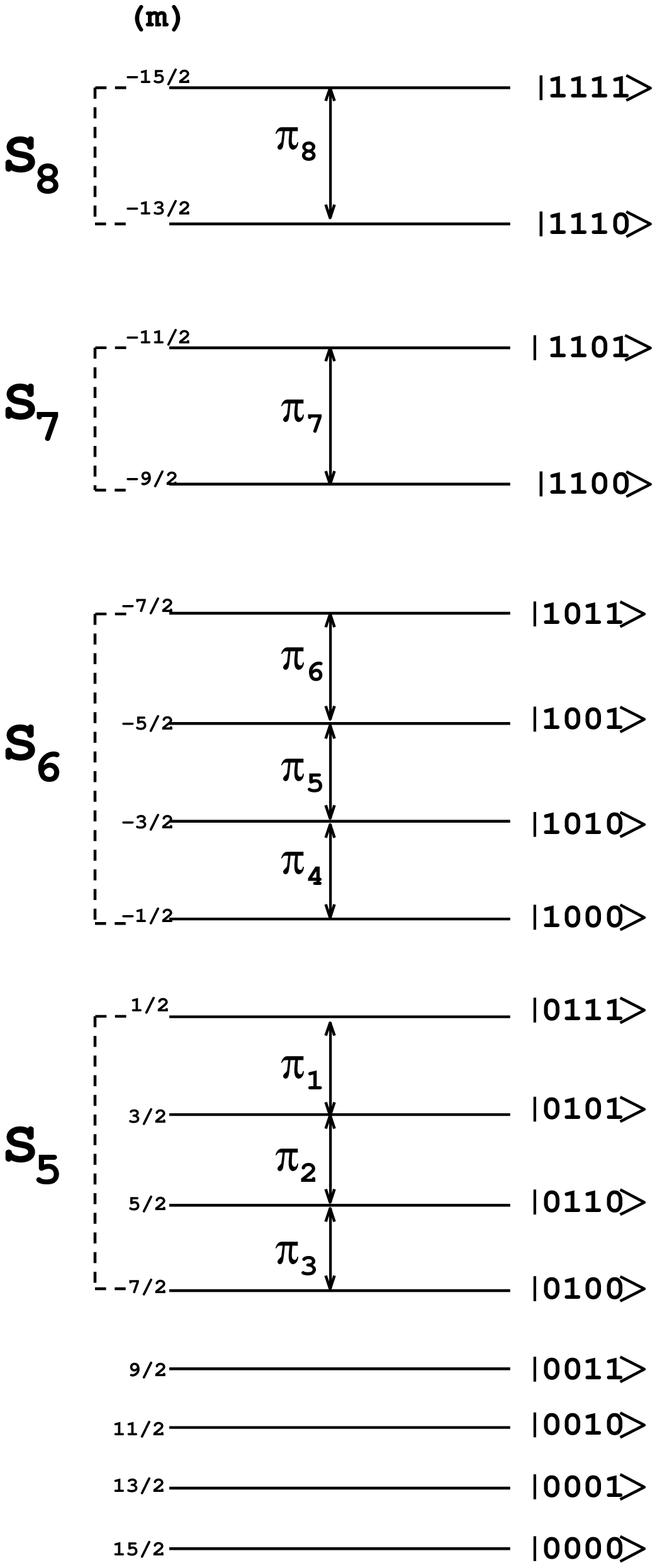,height=20cm}
\end{figure}
\hspace{3cm}
{\huge Figure 1}

\pagebreak
\vspace*{-3cm}
\begin{figure}
\setlength{\fboxrule}{.8pt}
\setlength{\fboxsep}{5pt}
\fbox{\epsfig{file=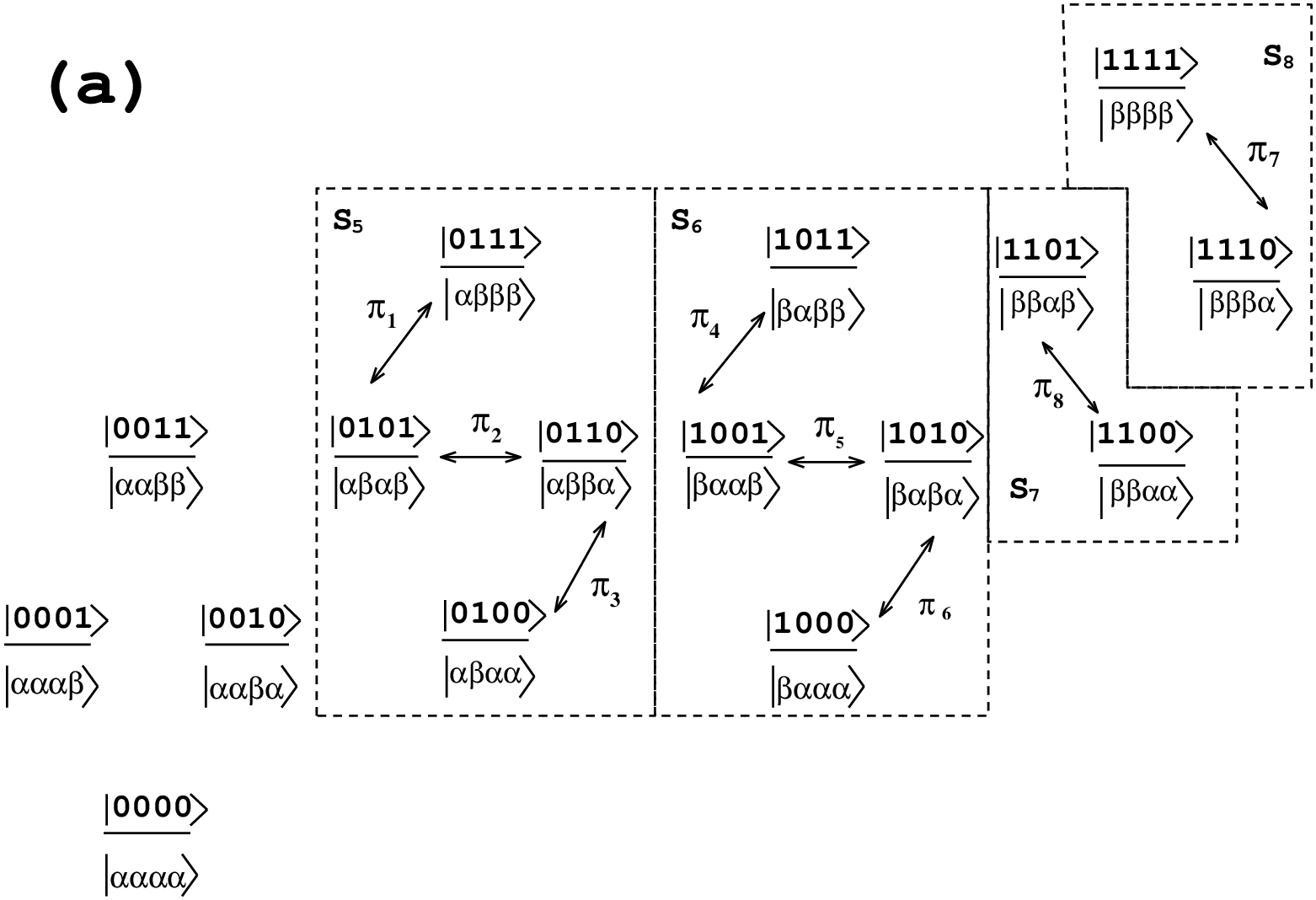,height=11cm,angle=270}}
\end{figure}
\vspace*{-1cm}
\begin{figure}
\setlength{\fboxrule}{.8pt}
\setlength{\fboxsep}{5pt}
\fbox{\epsfig{file=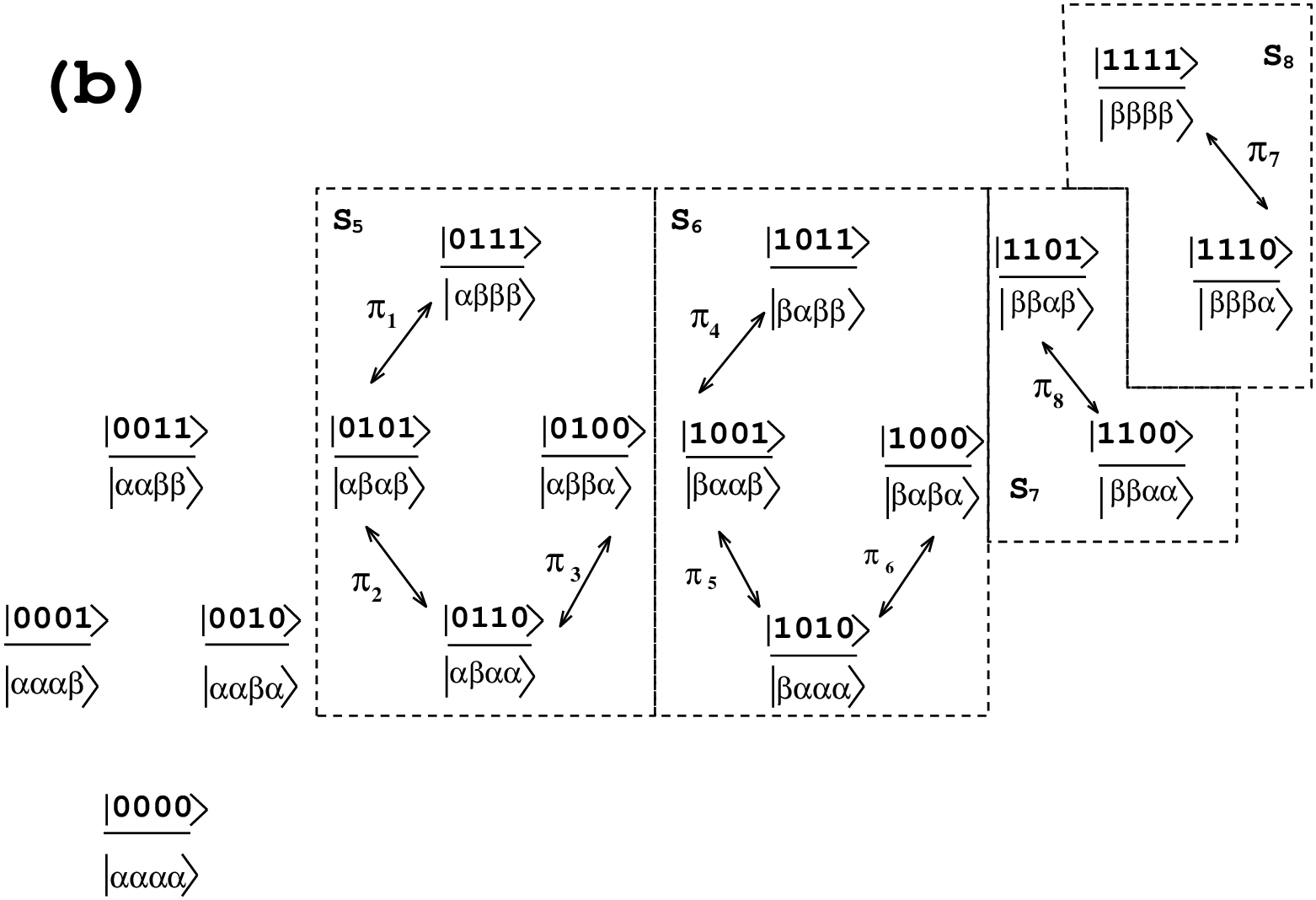,height=11cm,angle=270}}
\end{figure}
\vspace*{-1cm}
\begin{figure}
\setlength{\fboxrule}{.8pt}
\setlength{\fboxsep}{5pt}
\fbox{\epsfig{file=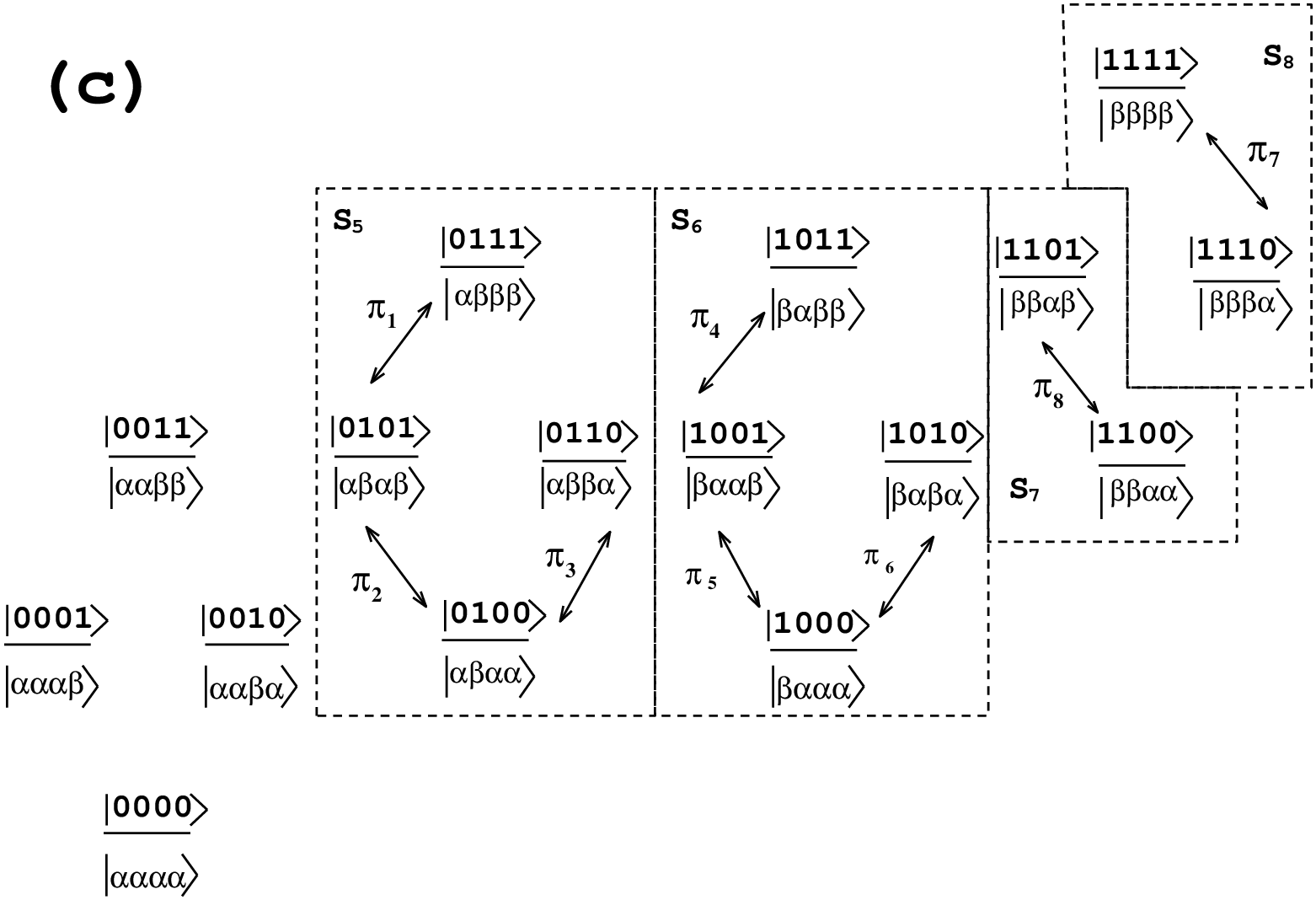,height=11cm,angle=270}}
\end{figure}
\hspace{6cm}
{\Large Figure 2}
\pagebreak
\begin{figure}
\epsfig{file=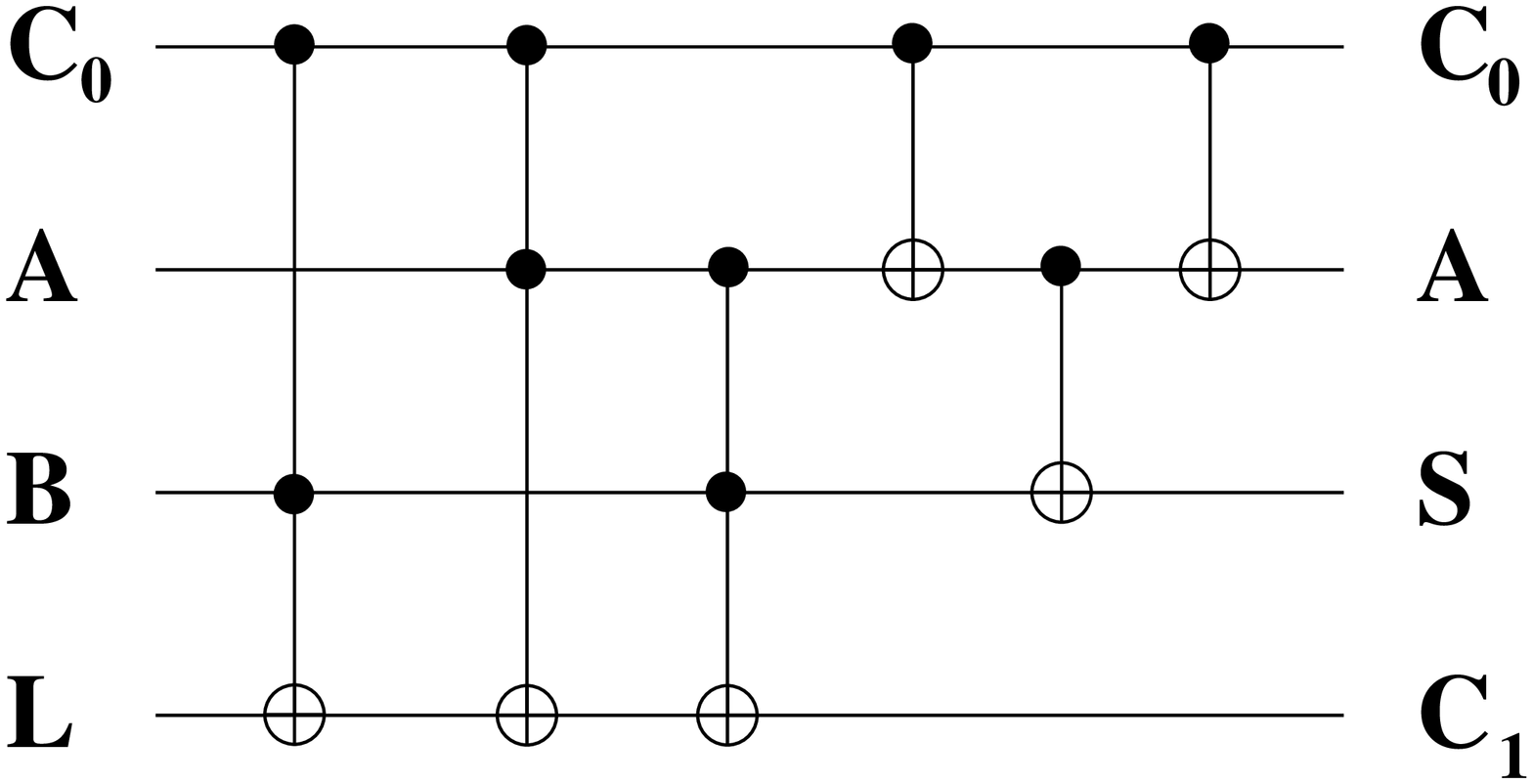,height=15cm,angle=270}
\end{figure}
\vspace{2cm}
\hspace{6cm}
{\huge Figure 3}
\pagebreak
\begin{figure}
\hspace{6cm}
\epsfig{file=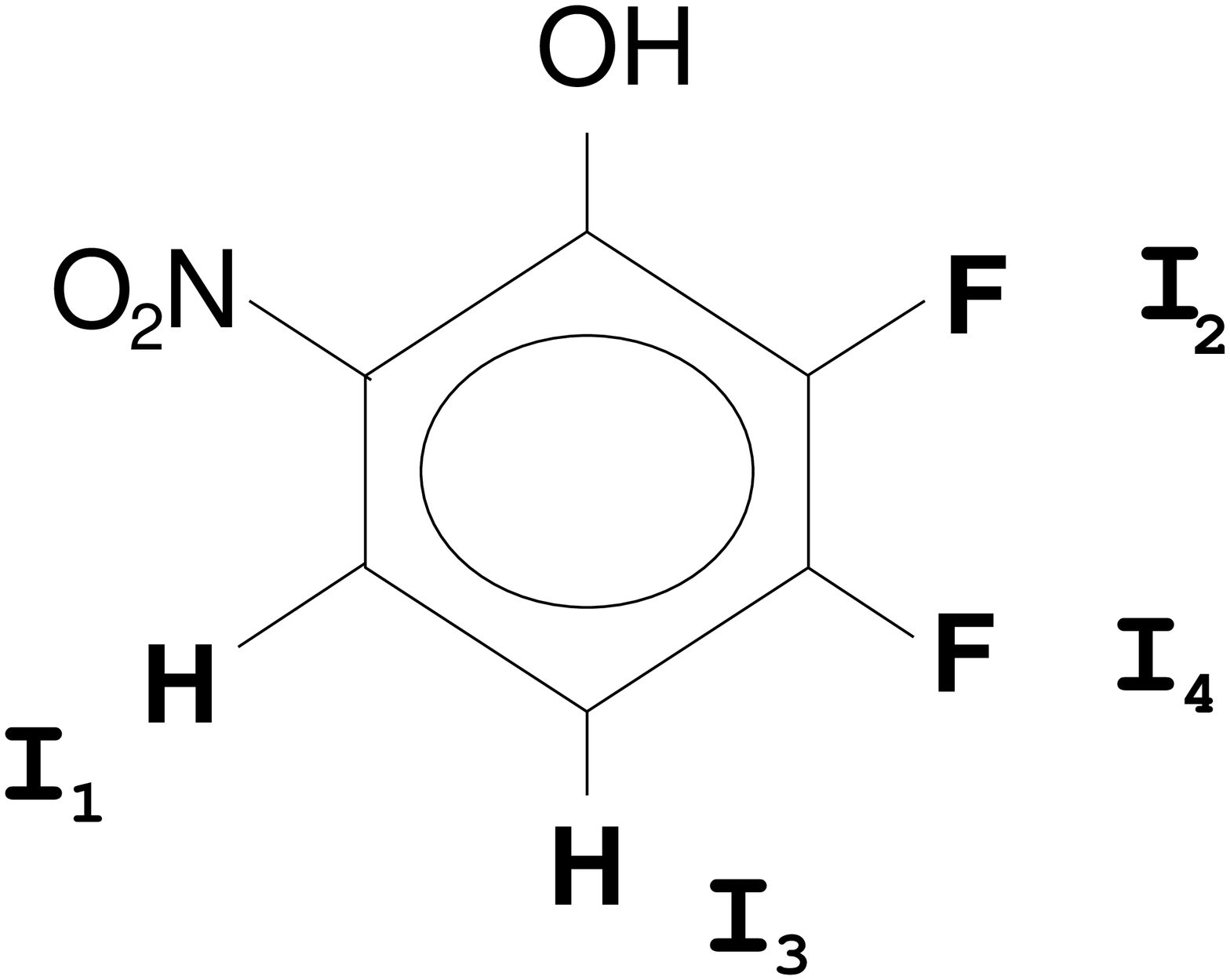,height=6cm,angle=270}
\end{figure}
\begin{figure}
\epsfig{file=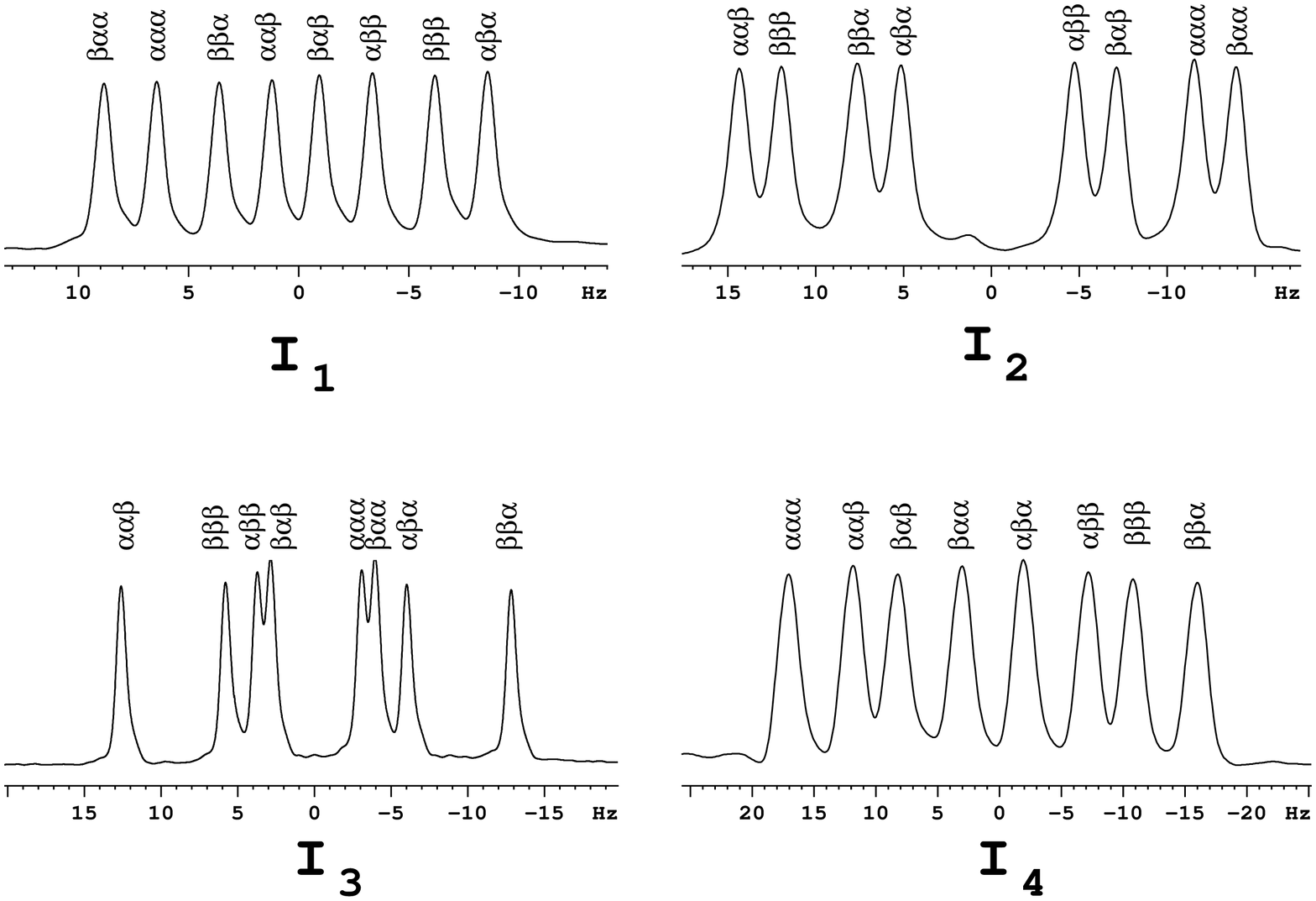,height=17cm,angle=270}
\end{figure}
\vspace{2cm}
\hspace{6cm}
{\huge Figure 4}
\pagebreak
\begin{figure}
\epsfig{file=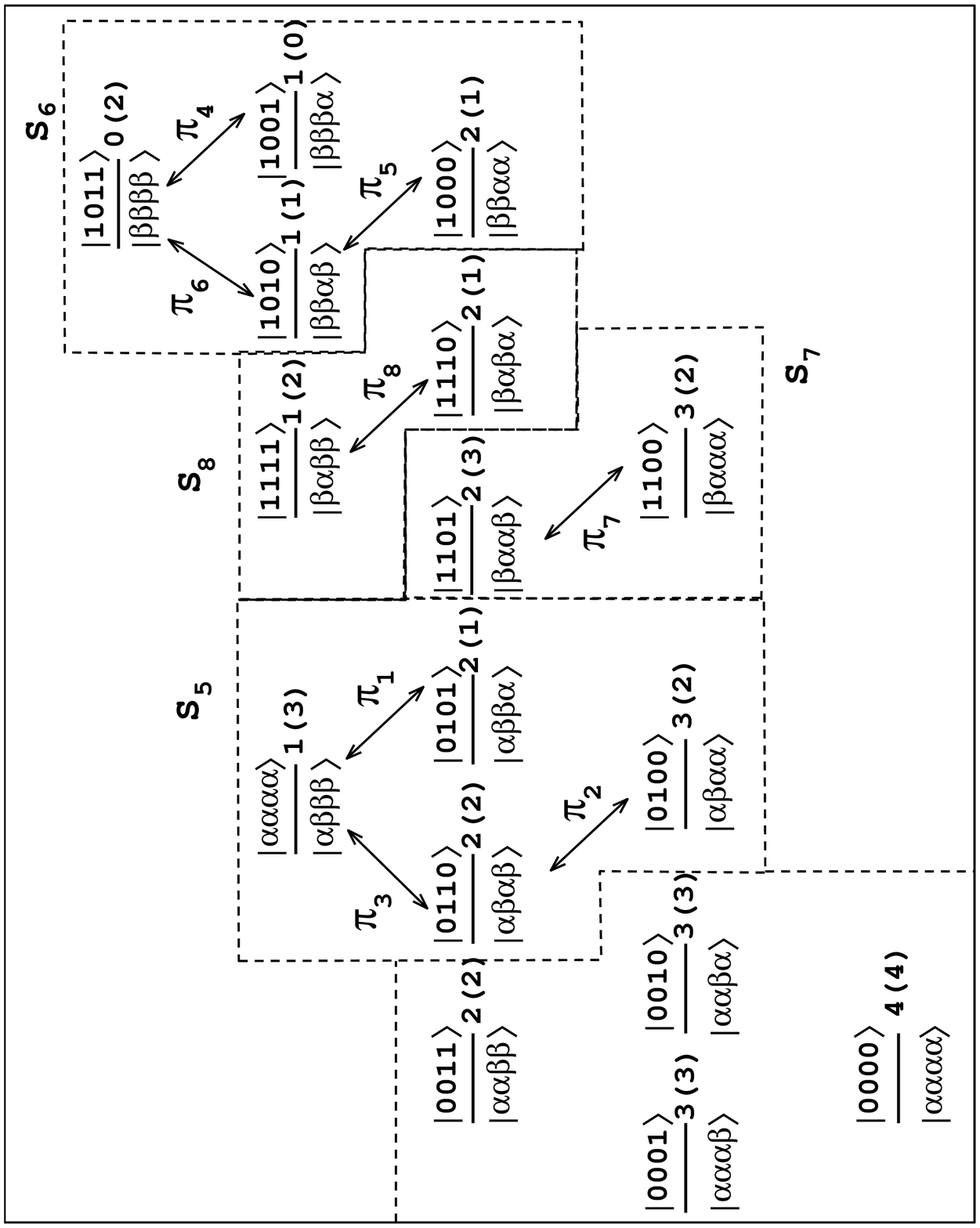,height=20cm}
\end{figure}
\hspace*{6cm}
{\huge Figure 5}
\pagebreak
\begin{figure}
\epsfig{file=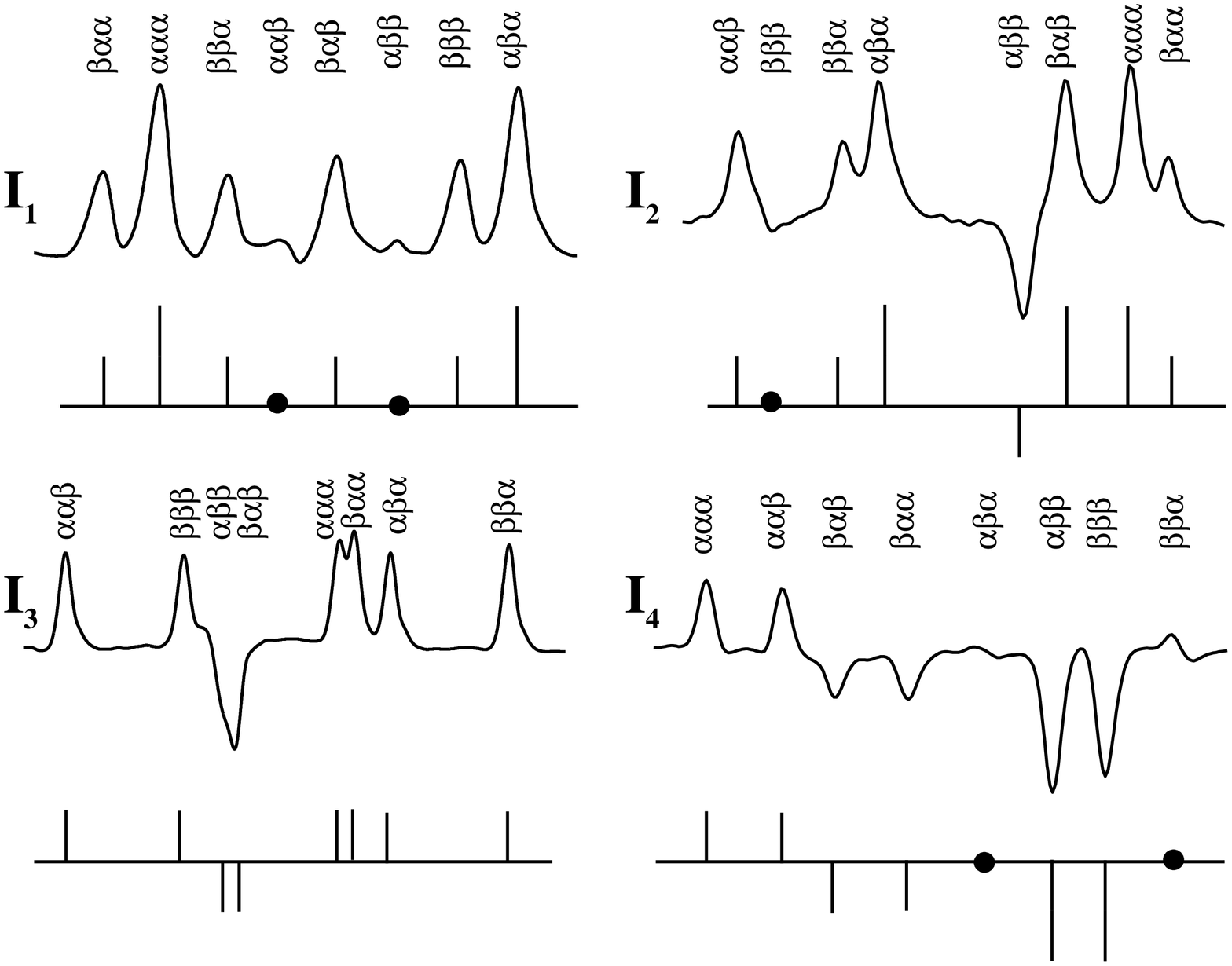,height=18cm,angle=270}
\end{figure}
\hspace*{6cm}
{\huge Figure 6}

\end{document}